\documentclass[aps,byrevtex,citeautoscript,showpacs,amsmath,amssymb]{revtex4}
\usepackage{graphicx}
\usepackage{dcolumn}
\usepackage{bm}
\usepackage{amsthm}
\usepackage{amsfonts,mathrsfs}
\usepackage{longtable}

\begin{document}

\title{Random Distance Distribution for Spherical Objects: 
General Theory and Applications to $n$-Dimensional Physics}

\author{Shu-Ju Tu}
\email[]{sjtu@physics.purdue.edu}
\author{Ephraim Fischbach}
\email[]{ephraim@physics.purdue.edu}
\affiliation{Department of Physics, Purdue University, West Lafayette, IN 47906}

\date{\today}

\begin{abstract}
A formalism is presented for analytically obtaining the probability
density function, \( P_{n}(s) \), for the random distance \( s \)
between two random points in an \( n \)-dimensional spherical object
of radius \( R \). Our formalism allows \( P_{n}(s) \) to be calculated
for a spherical \( n \)-ball having an arbitrary volume density,
and reproduces the well-known results for the case of uniform density.
The results find applications in stochastic geometry, computational
science, molecular biological systems, statistical physics, astrophysics,
condensed matter physics, nuclear physics, and elementary particle
physics. 
As one application of these results, 
we propose a new statistical method
obtained from our formalism to study random number generators in \( n\)-dimensions 
used in Monte Carlo simulations.
\end{abstract}


\maketitle

\section{Introduction}

In two recent papers ~\cite{Schleef,Parry_1}, geometric probability
techniques were developed to
calculate the functions \( P_{3}\left( s\right)  \)
which describe the probability density of finding a random distance
\( s \) separating two random points distributed in a uniform sphere
and in a uniform ellipsoid. As discussed in Refs.~\cite{Schleef,Parry_1,deltheil,hammersley,lord,Kendall,Santalo,Overhauser,Fischbach,Fischbach_k_t,sjtu,fedotov,jansons,mecke,miles},
these results are of interest as tools in mathematical physics,
and have numerous applications in other fields as well. 
Specifically, it was demonstrated in Refs.~\cite{Schleef,Parry_1,Fischbach,Fischbach_k_t,sjtu}
that knowing the random distance distribution in a spherical object greatly facilitates
the calculation of self-energies for spherical matter distributions
arising from electromagnetic, gravitational, or weak interactions. 

As an example we calculate the total electrostatic energy \( W_{3} \)
of a collection of \( Z \) charges uniformly distributed within the
same spherical volume of radius \( R \). For illustrative purposes,
we assume that \( Z \) is a large number. For each pair of charges the
potential energy due to the Coulomb interaction in Gaussian units is
\begin{equation}
\label{eq_new_00b}
V_{3}=\frac{e^{2}}{r_{12}}=\frac{e^{2}}{\left| \vec{r}_{2}-\vec{r}_{1}\right| },
\end{equation}
 where \( e \) is the elementary charge, \( \vec{r}_{1} \) (\( \vec{r}_{2} \)) is the
coordinate of the first (second) charge. 
The total Coulomb energy \( W_{3} \) can then
be expressed as\begin{eqnarray}
W_{3} & = & \frac{1}{2}\times \rho ^{2}\int r_{1}^{2}\, dr_{1}\int \sin \theta _{1}\, d\theta _{1}\int \, d\phi _{1}\int r_{2}^{2}\, dr_{2}\int \sin \theta _{2}\, d\theta _{2}\int \frac{1}{r_{12}}\, d\phi _{2}\nonumber \\
 & = & \frac{Z^{2}}{2}\times \frac{6}{5}\frac{e^{2}}{R},\label{eq_new_001} 
\end{eqnarray}
 where \( \rho =3Ze/4\pi R^{3} \)~\cite{jackson}. We note that Eq.~(\ref{eq_new_001})
requires evaluating a six-dimensional integral, 
and using the additional theorem for spherical harmonics~\cite{Arfken},
\begin{equation}
P_{l}\left( \cos \gamma \right) = \frac{4\pi }{2l+1}\sum _{m=-l}^{l}\left( -1\right) ^{m}Y^{m}_{l}\left( \theta _{1},\phi _{1}\right) Y_{l}^{-m}\left( \theta _{2},\phi _{2}\right) .
\end{equation}

Alternatively we can use the probability density function \( P_{3}\left( s\right) \)
giving the random distance distribution for a sphere with a uniform density    
to calculate \( W_{3} \). For a collection of \( Z \) charges 
there are \( Z\left( Z-1\right) /2 \) such pairs,
and hence the total Coulomb energy \( W_{3} \) is: 
\begin{equation}
\label{eq_new_002}
W_{3}=\frac{Z\left( Z-1\right) }{2}\times \int _{0}^{2R}P_{3}\left( s\right) \, \frac{e^{2}}{s}\, ds=\frac{Z\left( Z-1\right) }{2}\times \frac{6}{5}\frac{e^{2}}{R}\cong \frac{Z^{2}}{2}\times \frac{6}{5}\frac{e^{2}}{R},
\end{equation}
 where~\cite{deltheil} 
\begin{equation}
\label{eq_new_003}
P_{3}\left( s\right) =3\frac{s^{2}}{R^{3}}-\frac{9}{4}\frac{s^{3}}{R^{4}}+\frac{3}{16}\frac{s^{5}}{R^{6}}.
\end{equation}
We note that by using geometric probability techniques we can simplify the expression for \( W_{3} \)
from a six-dimensional non-trivial integral (Eq.~(\ref{eq_new_001}))
to a one-dimensional elementary integral (Eq.~(\ref{eq_new_002})).
Generalizing to \( n\)-dimensions, a calculation of the electrostatic energy \( W_{n}\) for a collection
of \( Z \) charges uniformly distributed within the same \( n \)-dimensional
spherical volume of radius \( R \), can be greatly simplified by utilizing 
the \( n \)-dimensional random distance distribution.
This reduces the complexity of calculating \( W_{n}\) from a \( 2n \)-dimensional
intractable integral involving \( n \)-dimensional spherical
harmonics, to a simple \( 1 \)-dimensional integral. 

The probability density function \( P_{n}(s) \) for the distribution
of the random distance \( s \) between two random points in
a uniform spherical \( n \)-ball 
is well known. Hence the object of the present paper is to generalize
the results of Refs.~\cite{Kendall,Santalo,Overhauser} to the case
of an arbitrarily non-uniform density distribution by using a new
method which we present below. Notice that the sample space \( \mathbf{B}_{n} \) for the
random points is a spherical \( n \)-ball of radius \( R \) defined as 
\begin{equation}
\label{eq_ball}
\mathbf{B}_{n}=\left\{ \left( x_{1},x_{2},\cdots ,x_{n}\right) \in \mathbf{R}^{n}:x_{1}^{2}+x_{2}^{2}+\cdots +x_{n}^{2}\leq R^{2}\right\} ,
\end{equation}
 where \begin{equation}
\label{eq_euclidean}
\mathbf{R}^{n}=\left\{ \boldsymbol {x}:\boldsymbol {x}=\left( x_{1},x_{2},\cdots ,x_{n}\right) \right\} 
\end{equation}
 represents \( n \)-dimensional Euclidean space.

To illustrate our formalism, we begin by deriving the probability
density function (PDF) for a spherical \( n \)-ball with a uniform
density distribution, and compare our results to those obtained earlier
by other means~\cite{Kendall,Santalo,Overhauser}. We then extend
this technique to a spherical \( n \)-ball with an arbitrary density
distribution, and this leads to a general-purpose master formula for
\( P_{n}\left( s\right)  \) given in Eq.~(\ref{eq_n_master_formula}).
The outline of this paper is as follows. In Sec.~\ref{sec_uniform_density}
we present our formalism and illustrate it by rederiving the well-known
results for a circle and for a sphere of uniform density. In Sec.~\ref{sec_spherical_density}
we extend this formalism to the case of non-uniform but spherically
symmetric density. In Sec.~\ref{sec_arbitrary_density} we develop
the formalism for the most general case of an arbitrary density distribution.
In Sec.~\ref{applications} we present some applications of our results.
These include the \( m \)th moment \( \left\langle s^{m}\right\rangle  \)
for a spherical space with a uniform and Gaussian density distribution,
the evaluation of the self-energy arising from \( \nu \bar{\nu } \)-exchange
interactions, the probability density functions for a sphere with
multiple shells that arises in neutron star model~\cite{Pines,Shapiro},
and finally a new proposed computational scheme for testing random
number generators in \( n \)-dimensions.

\section{Uniform Density Distributions\label{sec_uniform_density}}

\begin{figure}[t]
{\centering \resizebox*{2.50in}{!}{\includegraphics{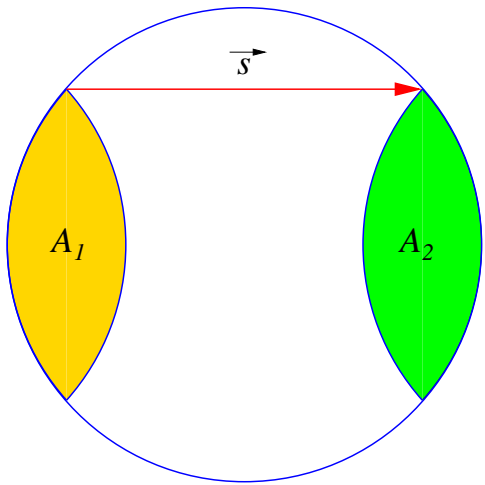}}\par}
\caption{Locus of points in a circle separated by a vector \protect\( \vec{s}=s\hat{x}\protect \).
For each random point \protect\( \vec{r}_{1}\protect \) in \protect\( A_{1}\protect \).
there is a unique random point \protect\( \vec{r}_{2}\protect \)
in \protect\( A_{2}\protect \) such that \protect\( \vec{s}=\vec{r}_{2}-\vec{r}_{1}\protect \).
\label{fig_001}}
\end{figure}

\begin{figure}[t]
{\centering \resizebox*{5.0in}{!}{\includegraphics{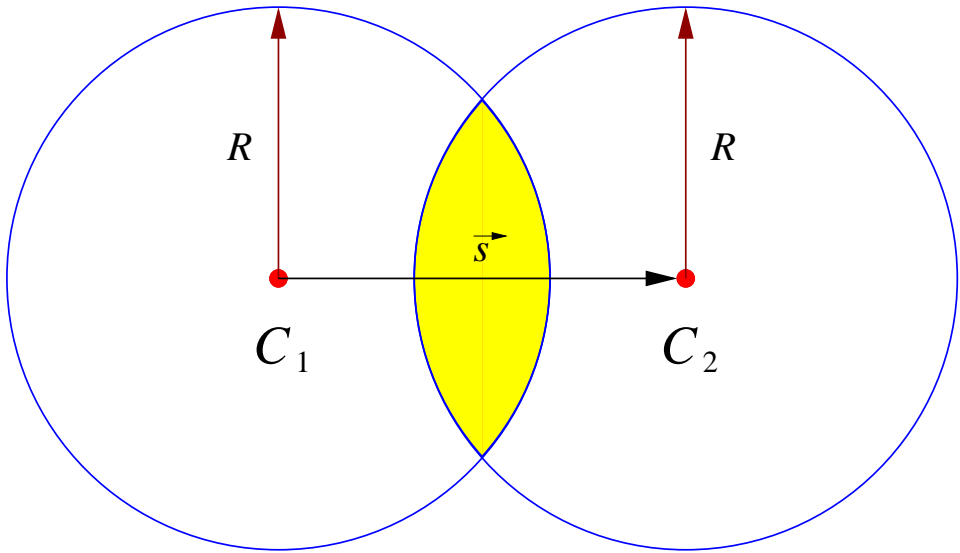}}\par}
\caption{The probability density that two random points are separated by a
random distance \protect\( \left| s\hat{x}\right| \protect \) in
a circle of radius \protect\( R\protect \) is proportional to the
shaded area given by the overlap of \protect\( C_{1}\protect \) and
\protect\( C_{2}\protect \), where \protect\( C_{1}\protect \) is
\protect\( x^{2}+y^{2}=R^{2}\protect \) and \protect\( C_{2}\protect \)
is \protect\( \left( x-s\right) ^{2}+y^{2}=R^{2}\protect \). \label{fig_002}}
\end{figure}

In this section we illustrate our formalism by deriving the PDF for
a circle of radius \( R \) having a spatially uniform density
characterized by a density function \( \rho  \), where \( \rho  \)
is an arbitrary constant. For two points randomly sampled inside the
circle located at \( \vec{r}_{1} \) and \( \vec{r}_{2} \) measured
from the center, define a random vector \( \vec{s}=\vec{r}_{2}-\vec{r}_{1} \)
and a random distance \( s=\left| \vec{s}\, \right|  \), where \( 0\leq s\leq 2R \).
To simplify the discussion, we translate the center of the circle
to the origin so that the equation for the circle is \( x^{2}+y^{2}=R^{2} \).
It is sufficient to initially consider those vectors \( \vec{s}\,  \)
which are aligned in the positive \( \hat{x} \) direction, since
rotational symmetry can eventually be used to extend our results to
those vectors \( \vec{s}\,  \) with arbitrary orientations. We begin
by identifying those pairs of points, \( \vec{r}_{1} \) and \( \vec{r}_{2} \),
which satisfy \( \vec{s}=s\hat{x} \). One set of random points for
\( \vec{r}_{1} \) is located in \( A_{1} \) and the other set of
random points for \( \vec{r}_{2} \) is located in \( A_{2} \) as
shown in Fig.~\ref{fig_001}. We observe that \( A_{2} \) is the
overlap area between the original circle \( C_{1} \) and an identical
circle \( C_{2} \) whose center is shifted from \( \left( 0,0\right)  \)
to \( \left( \left| \vec{s}\, \right| ,0\right)  \) as shown in Fig.~\ref{fig_002}.
Since the areas of \( A_{1} \) and \( A_{2} \) are equal, it follows
that the probability density of finding a given \( s=\left| s\hat{x}\right|  \)
in a circle of uniform density is proportional to the area of \( A_{2} \):
\begin{equation}
\int _{s-R}^{\frac{s}{2}}\, dx\int _{-\sqrt{R^{2}-\left( x-s\right) ^{2}}}^{\sqrt{R^{2}-\left( x-s\right) ^{2}}}\, dy+\int _{\frac{s}{2}}^{R}\, dx\int _{-\sqrt{R^{2}-x^{2}}}^{\sqrt{R^{2}-x^{2}}}\, dy.
\end{equation}
 Using rotational symmetry, this result can apply
to any orientation of \( \vec{s} \), where \( 0\leq \phi \leq 2\pi  \),
and hence the probability density \( P_{2}\left( s\right)  \) for
a circle of uniform density can be factorized as \begin{equation}
P_{2}\left( s\right) =2\pi s\times f\left( s\right) ,
\end{equation}
 where \( f\left( s\right)  \) is a function to be determined. If we impose
the normalization requirement 
\begin{equation}
\label{2_002}
\int _{0}^{2R}P_{2}(s)\, ds=1,
\end{equation}
 we then have  \begin{eqnarray}
P_{2}(s) & = & \frac{2\pi s\int _{\frac{s}{2}}^{R}\, dx\int _{-\sqrt{R^{2}-x^{2}}}^{\sqrt{R^{2}-x^{2}}}\, dy}{\int _{0}^{2R}\left( 2\pi s\int _{\frac{s}{2}}^{R}\, dx\int _{-\sqrt{R^{2}-x^{2}}}^{\sqrt{R^{2}-x^{2}}}\, dy\right) ds}\label{eq_2_002_a} \\
 & = & \frac{2s}{R^{2}}-\frac{s^{2}}{\pi R^{4}}\sqrt{4R^{2}-s^{2}}-\frac{4s}{\pi R^{2}}\sin ^{-1}\left( s/2R\right) .\label{eq_2_002_b} 
\end{eqnarray}
 Equation~(\ref{eq_2_002_b}) is identical to the results obtained
in Refs.~\cite{Kendall,Santalo} by other means.

The conclusion that emerges from this formalism is that the probability
density of finding two random points separated by a random vector
\( \vec{s}\,  \) in a circle of uniform density can be derived by
simply calculating the overlap region of that circle with an identical
circle obtained by shifting the center from the origin to \( \vec{s} \).
In the following discussion, we show that this result generalizes
to higher dimensions, and provides a simple way of calculating \( P_{n}(s) \)
for \( n\geq 3 \).

The above formalism for a circle of uniform density can be extended
to a sphere of uniform density. For a given \( s \), we select the
positive \( \hat{z} \) direction and study the distribution of the
random vectors \( \vec{s}\,  \) in this direction. It then follows
that the probability density of finding \( s\hat{z} \) is proportional
to \begin{equation}
\int _{s-R}^{\frac{s}{2}}\, dz\int _{-\sqrt{R^{2}-\left( z-s\right) ^{2}}}^{\sqrt{R^{2}-\left( z-s\right) ^{2}}}\, dx\int ^{\sqrt{R^{2}-\left( z-s\right) ^{2}-x^{2}}}_{-\sqrt{R^{2}-\left( z-s\right) ^{2}-x^{2}}}\, dy+\int _{\frac{s}{2}}^{R}\, dz\int _{-\sqrt{R^{2}-z^{2}}}^{\sqrt{R^{2}-z^{2}}}\, dx\int ^{\sqrt{R^{2}-z^{2}-x^{2}}}_{-\sqrt{R^{2}-z^{2}-x^{2}}}\, dy.
\end{equation}
 In a \( 3 \)-dimensional space, we note that the direction of the
random vector \( \vec{s}\,  \) can have the following range: \( 0\leq \theta \leq \pi  \)
and \( 0\leq \phi \leq 2\pi  \). Following the previous discussion,
we thus arrive at the following expression for a sphere with a uniform
density distribution: \begin{eqnarray}
P_{3}(s) & = & \frac{4\pi s^{2}\left( \int _{\frac{s}{2}}^{R}\, dz\int _{-\sqrt{R^{2}-z^{2}}}^{\sqrt{R^{2}-z^{2}}}\, dx\int ^{\sqrt{R^{2}-z^{2}-x^{2}}}_{-\sqrt{R^{2}-z^{2}-x^{2}}}\, dy\right) }{\int _{0}^{2R}4\pi s^{2}\left( \int _{\frac{s}{2}}^{R}\, dz\int _{-\sqrt{R^{2}-z^{2}}}^{\sqrt{R^{2}-z^{2}}}\, dx\int ^{\sqrt{R^{2}-z^{2}-x^{2}}}_{-\sqrt{R^{2}-z^{2}-x^{2}}}\, dy\right) ds}\label{eq_2_003_a} \\
 & = & 3\frac{s^{2}}{R^{3}}-\frac{9}{4}\frac{s^{3}}{R^{4}}+\frac{3}{16}\frac{s^{5}}{R^{6}}.\label{eq_2_003_b} 
\end{eqnarray}
 The result in Eq.~(\ref{eq_2_003_b}) agrees exactly with the expression
obtained previously in Refs.~\cite{Kendall,Santalo,Overhauser,deltheil,hammersley,lord}.

\begin{figure}[t]
{\centering\resizebox*{4.75in}{!}{\includegraphics{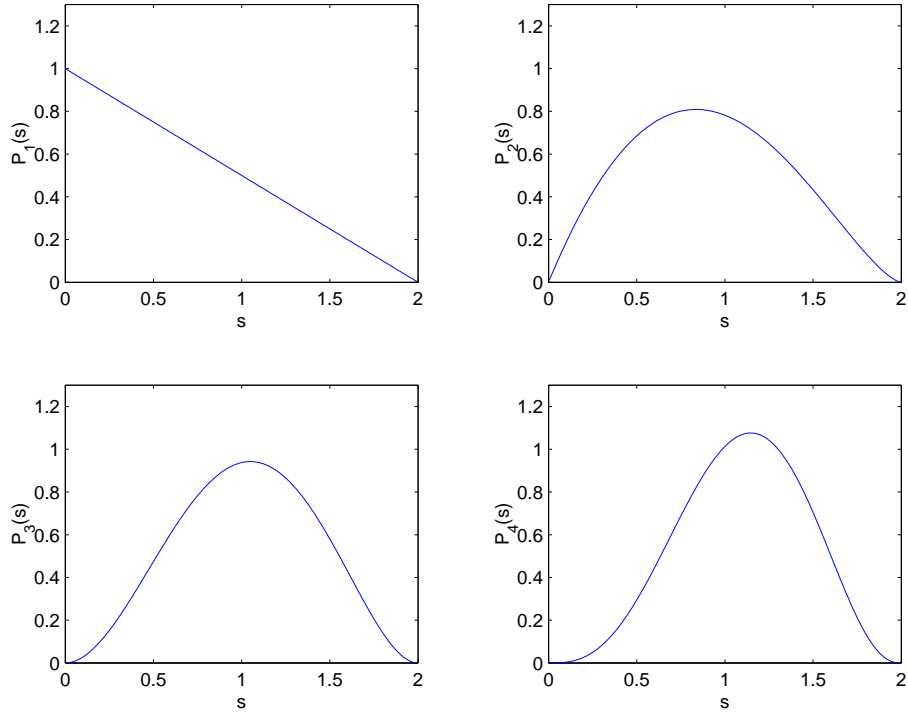}}\par} 
\caption{Plots of \protect\( P_{1}\left( s\right) \protect \), \protect\( P_{2}\left( s\right) \protect \),
\protect\( P_{3}\left( s\right) \protect \), and \protect\( P_{4}\left( s\right) \protect \)
for a uniform density distribution. 
Note that in all cases \protect\( \int_{0}^{2R}P_{n}(s)\,ds=1\protect\).
\label{fig_new_uniform}}
\end{figure}

The present formalism can be readily generalized to express \( P_{n}\left( s\right)  \)
for a sphere of uniform density in \(n\)-dimensions as 
\begin{equation}
\label{eq_2_100}
P_{n}(s)=\frac{s^{n-1}\int _{s/2}^{R}\, dx_{n}\int _{-\sqrt{R^{2}-x_{n}^{2}}}^{\sqrt{R^{2}-x_{n}^{2}}}\, dx_{1}\cdots \int _{-\sqrt{R^{2}-x_{n}^{2}-\cdots x_{n-2}^{2}}}^{\sqrt{R^{2}-x_{n}^{2}-\cdots x_{n-2}^{2}}}\, dx_{n-1}}{\int _{0}^{2R}\left( s^{n-1}\int _{s/2}^{R}\, dx_{n}\int _{-\sqrt{R^{2}-x_{n}^{2}}}^{\sqrt{R^{2}-x_{n}^{2}}}\, dx_{1}\cdots \int _{-\sqrt{R^{2}-x_{n}^{2}-\cdots x_{n-2}^{2}}}^{\sqrt{R^{2}-x_{n}^{2}-\cdots x_{n-2}^{2}}}\, dx_{n-1}\right) ds}.
\end{equation}
We find that if \( n \) is an even number, \begin{equation}
\label{eq_tu_even_uniform}
P_{n}(s)=\frac{n\times s^{n-1}}{R^{n}}\left[ \frac{2}{\pi }\cos ^{-1}\left( s/2R\right) -\frac{s}{\pi }\sum _{k=1}^{\frac{n}{2}}\frac{(n-2k)!!}{(n-2k+1)!!}\left( R^{2}-s^{2}/4\right) ^{\frac{n-2k+1}{2}}R^{2k-2-n}\right] ,
\end{equation}
 where \( 0!=0!!=1 \). If \( n \) is an odd number, \begin{equation}
\label{eq_tu_odd_uniform}
P_{n}(s)=\frac{n\times s^{n-1}}{R^{n}}\frac{n!!}{(n-1)!!}\sum _{k=0}^{\frac{n-1}{2}}\frac{(-1)^{k}}{2k+1}\frac{\left( \frac{n-1}{2}\right) !}{k!\left( \frac{n-1}{2}-k\right) !}\left[ 1-\left( s/2R\right) ^{2k+1}\right] .
\end{equation}
The functional forms of Eq.~(\ref{eq_2_100}) 
for \( n=1 \), \( 2 \), \( 3 \), and \( 4 \)
are shown in Fig.~\ref{fig_new_uniform}.

The cumulative distribution function (CDF) \cite{Feller,Gardiner,Kampen}
for \( P_{n}(s) \) is given by
\begin{eqnarray}
D_{n}(x) & = & \int _{0}^{x}P_{n}(s)\, ds\nonumber \\
 & = & \frac{x^{n}}{R^{n}}-\frac{B_{\alpha }\left( \frac{1}{2},\frac{n}{2}+\frac{1}{2}\right) }{B\left( \frac{1}{2},\frac{n}{2}+\frac{1}{2}\right) }\times \frac{x^{n}}{R^{n}}+2^{n}\times \frac{B_{\alpha }\left( \frac{n}{2}+\frac{1}{2},\frac{n}{2}+\frac{1}{2}\right) }{B\left( \frac{1}{2},\frac{n}{2}+\frac{1}{2}\right) },
\end{eqnarray}
where \( 0\leq x\leq 2R \), \( \alpha =x^{2}/4R^{2} \), and \( B_{\alpha } \)
is the incomplete beta function.

We summarize three important representations for the probability density
function \( P_{n}\left( s\right)  \) for a spherical \( n \)-ball
of radius \( R \) with a uniform density distribution as follows: 

\begin{enumerate}
\item \textsl{Integral representation}:\begin{equation}
\label{eq_new_integral}
P_{n}(s)=\frac{s^{n-1}\int _{\frac{s}{2}}^{R}\left( R^{2}-x^{2}\right) ^{\frac{n-1}{2}}\, dx}{\frac{1}{2n}B(\frac{n}{2}+\frac{1}{2},\frac{1}{2})R^{2n}}.
\end{equation}
 
\item \textsl{Generating function representation}:\begin{equation}
\label{eq_new_generating}
P_{n}(s)=\frac{s^{n-1}\frac{1}{n!}\left( \frac{\partial }{\partial h}\right) ^{n}_{h=0}\frac{1}{\sqrt{1-h^{2}}}\left[ \sin ^{-1}h-\sin ^{-1}\left( \frac{2h-\sqrt{4-s^{2}}}{2-h\sqrt{4-s^{2}}}\right) \right] }{\frac{1}{2n}B(\frac{n}{2}+\frac{1}{2},\frac{1}{2})},
\end{equation}
 where \( \left| h\right| <1 \) and \( R=1 \). We note that one
can obtain a number of identities and recursion relations for \( P_{n}\left( s\right)  \)
from the generating function representation in Eq.~(\ref{eq_new_generating}).
\item \textsl{Hypergeometric function representation}:\begin{equation}
\label{eq_new_hyper}
P_{n}(s)=\frac{2n}{B\left( \frac{n}{2}+\frac{1}{2},\frac{1}{2}\right) }\times \left[ \, _{2}F_{1}\left( a,b,c,\alpha \right) R-\frac{s}{2}\, \, _{2}F_{1}\left( a,b,c,\beta \right) \right] \times \frac{s^{n-1}}{R^{n+1}},
\end{equation}
 where \( a=1/2 \), \( b=1/2-n/2 \), \( c=3/2 \), \( \alpha =1 \),
\( \beta =s^{2}/4R^{2} \), and \( \, _{2}F_{1}\left( \cdots \right)  \)
is the hypergeometric function~\cite{Arfken}. Notice that we can
obtain the orthogonality relations for \( P_{n}\left( s\right)  \)
from the hypergeometric function representation as shown in Eq.~(\ref{eq_new_hyper}). 
\end{enumerate}
In Ref.~\cite{sjtu} additional identities and recursion relations
for \( P_{n}\left( s\right)  \) are discussed in greater detail.

\section{Spherically Symmetric Density Distributions\label{sec_spherical_density}}

In this section we extend the previous results to the case of a circle
with a variable (but spherically symmetric) density characterized
by a density function \( \rho \left( r\right)  \). Following the
derivation presented in the previous section, we note that for any
random vector \( \vec{s}=s\hat{x} \), if the second random point
\( \vec{r}_{2} \) carries the density information \( \rho \left( x,y\right)  \),
then the first random point \( \vec{r}_{1} \) should have the density
information \( \rho \left( x-s,y\right) . \) It then follows \( P_{2}(s) \)
can be expressed as \begin{eqnarray}
P_{2}(s) & = & \frac{s\int _{\frac{s}{2}}^{R}\, dx\int _{-\sqrt{R^{2}-x^{2}}}^{\sqrt{R^{2}-x^{2}}}\rho \left( x-s,y\right) \times \rho \left( x,y\right) \, dy}{\int _{0}^{2R}\left( s\int _{\frac{s}{2}}^{R}\, dx\int _{-\sqrt{R^{2}-x^{2}}}^{\sqrt{R^{2}-x^{2}}}\rho \left( x-s,y\right) \times \rho \left( x,y\right) \, dy\right) ds}.\label{eq_non_uniform_2} 
\end{eqnarray}
A general formula for \( P_{n}\left( s\right)  \) for an \( n \)-dimensional
spherical ball of radius \( R \) having a spherically symmetric density
can be derived, and we find \begin{equation}
\label{eq_spherical_001}
P_{n}(s)=\frac{s^{n-1}\int _{\frac{s}{2}}^{R}\, dx_{n}\int _{-\sqrt{R^{2}-x_{n}^{2}}}^{\sqrt{R^{2}-x_{n}^{2}}}\, dx_{1}\cdots \int _{-\sqrt{R^{2}-x_{n}^{2}-x_{1}^{2}-\cdots x_{n-1}^{2}}}^{\sqrt{R^{2}-x_{n}^{2}-x_{1}^{2}-\cdots x_{n-1}^{2}}}\rho _{1}\times \rho _{2}\, dx_{n-1}}{\int _{0}^{2R}\left( s^{n-1}\int _{\frac{s}{2}}^{R}\, dx_{n}\int _{-\sqrt{R^{2}-x_{n}^{2}}}^{\sqrt{R^{2}-x_{n}^{2}}}\, dx_{1}\cdots \int _{-\sqrt{R^{2}-x_{n}^{2}-x_{1}^{2}-\cdots x_{n-1}^{2}}}^{\sqrt{R^{2}-x_{n}^{2}-x_{1}^{2}-\cdots x_{n-1}^{2}}}\rho _{1}\times \rho _{2}\, dx_{n-1}\right) ds},
\end{equation}
 where\begin{eqnarray}
\rho _{1} & = & \rho \left( x_{1},x_{2},\cdots ,x_{n}-s\right) ,\nonumber \\
\rho _{2} & = & \rho \left( x_{1},x_{2},\cdots ,x_{n}\right) .
\end{eqnarray}
 Some analytical results for a sphere with various spherically symmetric
density distributions can be found in Ref.~\cite{sjtu}.

\begin{figure}[t]
{\centering \resizebox*{4.75in}{!}{\includegraphics{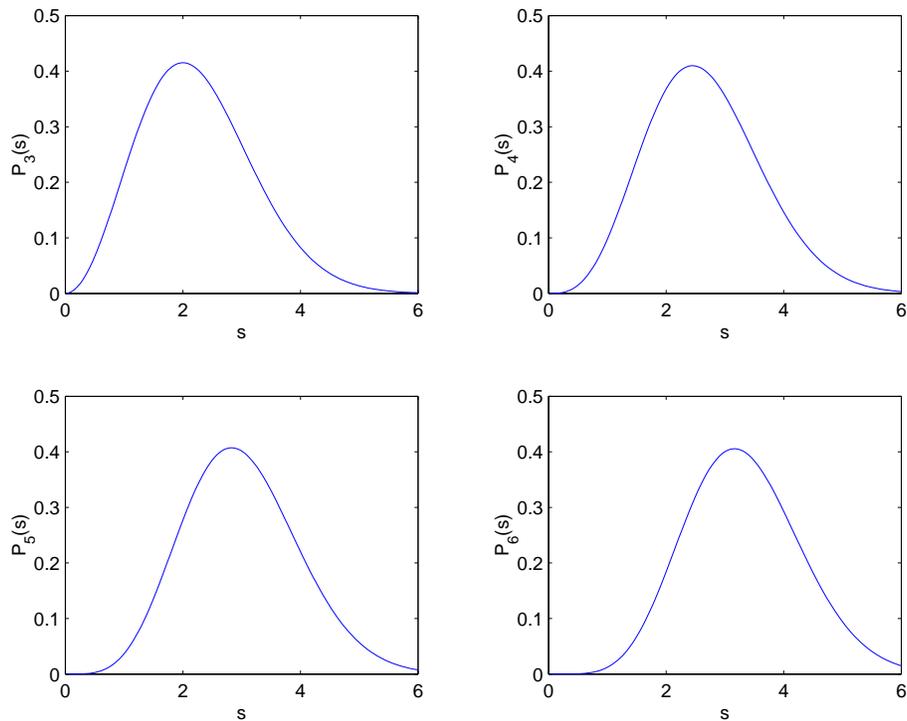}} \par}
\caption{Plots of \protect\( P_{3}\left( s\right) \protect \), \protect\( P_{4}\left( s\right) \protect \),
\protect\( P_{5}\left( s\right) \protect \), and \protect\( P_{6}\left( s\right) \protect \)
for a Gaussian density distribution. 
Note that in all cases the functions are normalized such that 
\protect\( \int_{0}^{\infty}P_{n}(s)\, ds=1\protect\).
\label{fig_new_gaussian}}
\end{figure}

As an application of Eq.~(\ref{eq_spherical_001}) we consider the
case of an \( n \)-dimensional spherical space of radius \( R\rightarrow \infty  \)
with a Gaussian density given by \begin{equation}
\label{eq_n_gaussian_density}
\rho _{n}(r)=\frac{N}{(2\pi )^{\frac{n}{2}}\sigma ^{n}}e^{-\frac{1}{2}\frac{r^{2}}{\sigma ^{2}}},
\end{equation}
 where\begin{equation}
\label{eq_normalization_gaussian}
N=\lim _{R\rightarrow \infty }n\frac{\pi ^{\frac{n}{2}}}{\Gamma \left( \frac{n}{2}+1\right) }\int _{0}^{R}\rho _{n}(r)r^{n-1}dr.
\end{equation}
 In Eq.~(\ref{eq_normalization_gaussian}) \( r \) is measured from
the center, and the integral is over all space. The PDF for an \( n \)-dimensional
spherical Gaussian space can then be obtained as
\begin{equation}
\label{eq_n_pdf_gaussian}
P_{n}(s)=\frac{s^{n-1}e^{-\frac{s^{2}}{4\sigma ^{2}}}}{2^{n-1}\Gamma \left( \frac{n}{2}\right) \sigma ^{n}}.
\end{equation}
 Fig.~\ref{fig_new_gaussian} displays the functions in Eq.~(\ref{eq_n_pdf_gaussian})
for \( n=3 \), \( 4 \), \( 5 \), and \( 6 \), where \( \sigma =1 \).
Finally, we note that the maximum probability density, denoted by
\( s_{\mathrm{max}} \), occurs at \begin{equation}
s_{\mathrm{max}}=\sqrt{2(n-1)}\, \sigma .
\end{equation}

\section{Arbitrary Density Distributions\label{sec_arbitrary_density}}

We consider in this section the probability density functions for
a spherical \( n \)-ball having an arbitrary density characterized
by a density function \( \rho  \). We begin with a circle and use
the conventional notation for polar coordinates: \( x=r\cos \phi  \)
and \( y=r\sin \phi  \). In a \( 2 \)-dimensional space, the random
vector \( \vec{s}\,  \) can be characterized by an angle \( \phi  \)
in the range \( 0\leq \phi \leq 2\pi  \). Associate each random unit
vector \( \hat{s}\left( \phi \right)  \) with a rotation operator
\( \mathscr{R} \) such that\begin{equation}
\left| \hat{x}\right\rangle =\mathscr{R}\left| \hat{s}\left( \phi \right) \right\rangle .
\end{equation}
 To ensure that the product of \( \rho \left( \vec{r}_{1}\right)  \)
and \( \rho \left( \vec{r}_{2}\right)  \) maintains the correct density
information, we use a \( 2\times 2 \) matrix \begin{equation}
R_{2\times 2}(\phi )=\left[ \begin{array}{cc}
\cos \phi  & -\sin \phi \\
\sin \phi  & \cos \phi 
\end{array}\right] 
\end{equation}
 to characterize this particular operator \( \mathscr{R} \) such that
\begin{equation}
\rho (x,y)\longrightarrow \rho \left( \cos \phi \, x-\sin \phi \, y,\sin \phi \, x+\cos \phi \, y\right) .
\end{equation}
Notice that \( R_{2\times 2}(\phi ) \) is an orthogonal matrix which
satisfies \( R_{2\times 2}^{-1}(\phi )=R_{2\times 2}^{T}(\phi ) \)
and its determinant is \( +1 \), where \( T \) denotes the transpose. 

\begin{figure}[t]
{\centering \resizebox*{4.75in}{!}{\includegraphics{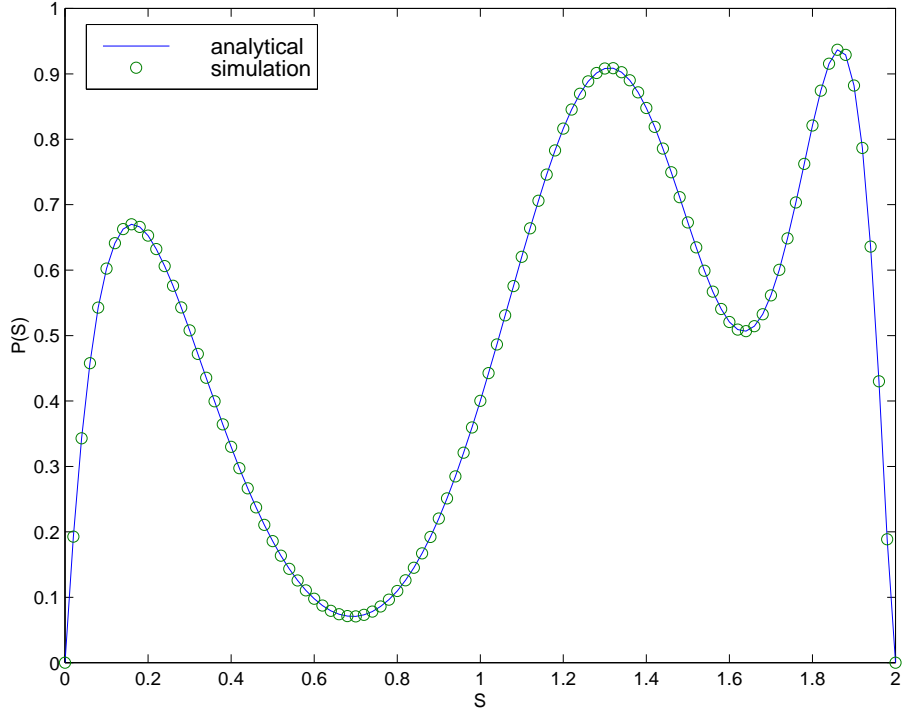}} \par}
\caption{Plot of \protect\( P_{2}(s)\protect \) as a function of \protect\( s\protect \)
for the case \protect\( R=1\protect \) and \protect\( \rho \left( x,y\right) \propto x^{4}y^{4}\protect \).\label{density_2}}
\end{figure}

We can then express \( P_{2}\left( s\right)  \) for a circle with
an arbitrary density distribution as\begin{equation}
P_{2}(s)=\frac{s\int _{0}^{2\pi }\, d\phi \int _{s/2}^{R}\, dx\int _{-\sqrt{R^{2}-x^{2}}}^{\sqrt{R^{2}-x^{2}}}\rho \left( x',y'\right) \times \rho \left( x'',y''\right) \, dy}{\int _{0}^{2R}\left( s\int _{0}^{2\pi }\, d\phi \int _{s/2}^{R}\, dx\int _{-\sqrt{R^{2}-x^{2}}}^{\sqrt{R^{2}-x^{2}}}\rho \left( x',y'\right) \times \rho \left( x'',y''\right) \, dy\right) ds}
\end{equation}
 where\begin{eqnarray*}
\rho \left( x',y'\right)  & = & \rho \left( \cos \phi \, (x-s)-\sin \phi \, y,\, \sin \phi \, (x-s)+\cos \phi \, y\right) ,\\
\rho \left( x'',y''\right)  & = & \rho \left( \cos \phi \, x-\sin \phi \, y,\, \sin \phi \, x+\cos \phi \, y\right) .
\end{eqnarray*}
 Figure~\ref{density_2} exhibits \( P_{2}(s) \) when \( R=1 \),
and illustrates the agreement between the Monte Carlo simulation
and the analytical result. 

The preceding discussion can be generalized to the case of a spherical
\( n \)-ball with an arbitrary density. Define the following hyperspherical
coordinates~\cite{fleming},\begin{eqnarray}
x_{1} & = & r\sin \theta _{n-2}\sin \theta _{n-3}\cdots \cdots \cdots \cdots \cdots \cdots \sin \theta _{2}\sin \theta _{1}\cos \phi ,\nonumber \\
x_{2} & = & r\sin \theta _{n-2}\sin \theta _{n-3}\cdots \cdots \cdots \cdots \cdots \cdots \sin \theta _{2}\sin \theta _{1}\sin \phi ,\nonumber \\
 & \vdots  & \nonumber \\
x_{i} & = & r\sin \theta _{n-2}\sin \theta _{n-3}\cdots \cdots \sin \theta _{i-1}\cos \theta _{i-2},\nonumber \\
 & \vdots  & \nonumber \\
x_{n-1} & = & r\sin \theta _{n-2}\cos \theta _{n-3},\nonumber \\
x_{n} & = & r\cos \theta _{n-2}.
\end{eqnarray}
 where \( \theta _{i}\in [0,\pi ] \) and \( \phi \in [0,2\pi ] \).
Associate each random unit vector \( \hat{s}\left( \theta _{1},\cdots ,\phi \right)  \)
with a rotation operator \( \mathscr{R} \) such that\begin{equation}
\label{eq_4_102}
\left| \hat{x}_{n}\right\rangle =\mathscr{R}\left| \hat{s}\left( \theta _{1},\cdots ,\phi \right) \right\rangle .
\end{equation}
 The \( n\times n \) matrix representation for the rotation operator
\( \mathscr{R} \) in Eq.~(\ref{eq_4_102}) can be expressed as \begin{equation}
\label{eq_n_rotation_operator}
R_{n\times n}\left( \theta _{n-2},\cdots ,\theta _{1},\phi \right) =R_{n\times n}(\phi )\times R_{n\times n}(\theta _{1})\times \cdots \times R_{n\times n}(\theta _{n-2}),
\end{equation}
 where \( n\geq 3 \). The various matrices on the right-hand side
of Eq.~(\ref{eq_n_rotation_operator}) are defined as follows:

\begin{enumerate}
\item \( R_{n\times n}(\phi ) \): The matrix elements are \( a_{11}=\cos \phi  \),
\( a_{12}=-\sin \phi  \), \( a_{21}=\sin \phi  \), \( a_{22}=\cos \phi  \),
for \( i\in [1,2] \) and \( j\in [3,n] \) \( a_{ij}=0 \), for \( i\in [3,n] \)
and \( i\in [1,2] \) \( a_{ij}=0 \), for \( i\in [3,n] \) and \( j\in [3,n] \)
\( a_{ij}=\delta _{ij} \).
\item \( R_{n\times n}\left( \theta _{1}\right)  \): The matrix elements
are \( a_{11}=\cos \theta _{1} \), \( a_{12}=0 \), \( a_{13}=\sin \theta _{1} \),
\( a_{21}=0 \), \( a_{22}=1 \), \( a_{23}=0 \), \( a_{31}=-\sin \theta _{1} \),
\( a_{32}=0 \), \( a_{33}=\cos \theta _{1} \), for \( i\in [1,3] \)
and \( j\in [4,n] \) \( a_{ij}=0 \), for \( i\in [4,n] \) and \( j\in [1,3] \)
\( a_{ij}=0 \), for \( i\in [4,n] \) and \( j\in [4,n] \) \( a_{ij}=\delta _{ij} \).
\item \( R_{n\times n}\left( \theta _{k}\right)  \) and \( k\in [2,n-3] \):
The matrix elements are \( a_{k+1,k+1}=\cos \theta _{k} \), \( a_{k+1,k+2}=\sin \theta _{k} \),
\( a_{k+2,k+1}=-\sin \theta _{k} \), \( a_{k+2,k+2}=\cos \theta _{k} \),
for \( i\in [1,k] \) and \( j\in [1,k] \) \( a_{ij}=\delta _{ij} \),
for \( i\in [k+3,n] \) and \( j\in [k+3,n] \) \( a_{ij}=\delta _{ij} \),
for \( i\in [1,k] \) and \( j\in [k+1,n] \) \( a_{ij}=0 \), for
\( i\in [k+1,n] \) and \( j\in [1,k] \) \( a_{ij}=0 \), for \( i\in [k+1,k+2] \)
and \( j\in [k+3,n] \) \( a_{ij}=0 \), for \( i\in [k+3,n] \) and
\( j\in [k+1,k+2] \) \( a_{ij}=0 \).
\item \( R_{n\times n}\left( \theta _{n-2}\right)  \): The matrix elements
are \( a_{n-1,n-1}=\cos \theta _{n-2} \), \( a_{n-1,n}=\sin \theta _{n-2} \),
\( a_{n,n-1}=-\sin \theta _{n-2} \), \( a_{n,n}=\cos \theta _{n-2} \),
for \( i\in [1,n-2] \) and \( j\in [1,n-2] \) \( a_{ij}=\delta _{ij} \),
for \( i\in [1,n-2] \) and \( j\in [n-1,n] \) \( a_{ij}=0 \), for
\( i\in [n-1,n] \) and \( j\in [1,n-2] \) \( a_{ij}=0 \). 
\end{enumerate}
Notice that all the matrices in Eq.~(\ref{eq_n_rotation_operator})
are orthogonal and their determinants are \( +1 \). The matrix elements
for \( R_{n\times n}(\theta _{n-2},\cdots ,\theta _{1},\phi ) \)
can be summarized as follows:

\begin{enumerate}
\item The matrix elements for the \( 1 \)st column are \( a_{11}=\cos \theta _{1}\cos \phi  \),
\( a_{21}=\cos \theta _{1}\sin \phi  \), \( a_{31}=-\sin \theta _{1} \),
and for \( i\in [4,n] \) \( a_{i1}=0 \).
\item The matrix elements for the \( 2 \)nd column are \( a_{12}=-\sin \phi  \),
\( a_{22}=\cos \phi  \), and for \( i\in [3,n] \) \( a_{i2}=0 \).
\item The matrix elements for the \( j \)th column where \( j\in [3,n-1] \)
are \( a_{j+1,j}=-\sin \theta _{j-1} \), for \( i\in [j+2,n] \)
\( a_{ij}=0 \), and for \( i\in [1,j] \) \( a_{ij}=\cos \theta _{j-1}\times x_{i} \),
where \( x_{i}=\hat{r}\times \hat{x}_{i} \) and \( x_{i} \) is the
\( i \)th Cartesian coordinate component for a unit vector \( \hat{r} \)
in \( j \)-dimensions. For example:\begin{eqnarray*}
a_{14} & = & \cos \theta _{3}\sin \theta _{2}\sin \theta _{1}\cos \phi ,\\
a_{24} & = & \cos \theta _{3}\sin \theta _{2}\sin \theta _{1}\sin \phi ,\\
a_{34} & = & \cos \theta _{3}\sin \theta _{2}\cos \theta _{1},\\
a_{44} & = & \cos \theta _{3}\cos \theta _{2}.
\end{eqnarray*}

\item The matrix elements for the \( n \)th column are for \( i\in [1,n] \)
\( a_{in}=x_{i} \), where \( x_{i}=\hat{r}\times \hat{x}_{i} \)
and \( x_{i} \) is the \( i \)th Cartesian coordinate component
for a unit vector \( \hat{r} \) in \( n \)-dimensions. 
\end{enumerate}
Additionally, it is convenient to define the following transforms:\begin{eqnarray}
x_{i}' & = & \sum _{j=1}^{n}a_{ij}\left( x_{j}-\delta _{jn}s\right) ,\\
x_{i}'' & = & \sum _{j=1}^{n}a_{ij}x_{j},
\end{eqnarray}
 where \( a_{ij} \) are the matrix elements for \( R_{n\times n}\left( \theta _{n-2},\cdots ,\theta _{1},\phi \right)  \)
defined in Eq.~(\ref{eq_n_rotation_operator}). The master equation
of \( P_{n}(s) \) for a spherical \( n \)-ball with an arbitrary
density characterized by a density function \( \rho \left( \boldsymbol {x}\right) =\rho \left( x_{1},x_{2},\cdots ,x_{n}\right)  \)
can then be formulated as
\begin{equation}
\label{eq_n_master_formula}
P_{n}(s)=\frac{s^{n-1}\times \mathbf{F}_{n}\left( x,\theta \right) }{\int _{0}^{2R}\left[ s^{n-1}\times \mathbf{F}_{n}\left( x,\theta \right) \right] \, ds},
\end{equation}
 where
\begin{eqnarray*}
\mathbf{F}_{n}\left( x,\theta \right)  & = & \int _{0}^{\pi }\sin ^{n-2}\theta _{n-2}\, d\theta _{n-2}\int _{0}^{\pi }\sin ^{n-3}\theta _{n-3}\, d\theta _{n-3}\cdots \nonumber \\
 &  & \cdots \cdots \int _{0}^{\pi }\sin \theta _{1}\, d\theta _{1}\int _{0}^{2\pi }\, d\phi \int _{s/2}^{R}\, dx_{n}\int _{-\sqrt{R^{2}-x_{n}^{2}}}^{\sqrt{R^{2}-x_{n}^{2}}}\, dx_{1}\cdots \nonumber \\
 &  & \cdots \cdots \cdots \int ^{\sqrt{R^{2}-x_{n}^{2}-\cdots x_{n-2}^{2}}}_{-\sqrt{R^{2}-x_{n}^{2}-\cdots x_{n-2}^{2}}}\, \rho \left( \boldsymbol {x}'\right) \times \rho \left( \boldsymbol {x}''\right) \, dx_{n-1},
\end{eqnarray*}
 and\begin{eqnarray*}
\rho \left( \boldsymbol {x}'\right)  & = & \rho \left( x_{1}',x_{2}',\cdots ,x_{n}'\right) ,\\
\rho \left( \boldsymbol {x}''\right)  & = & \rho \left( x''_{1},x''_{2},\cdots ,x''_{n}\right) .
\end{eqnarray*}

The probability density function for the random distance distribution
including Euclidean distance and geodesic distance for an \( n \)-dimensional
sphere (i.e. the boundary of a spherical \( n\)-ball) 
with an arbitrary surface density distribution will be discussed in Ref.~\cite{sjtu4}.

\section{Applications\label{applications}}

\subsection{\protect\( m\protect \)th Moment of the Distance}

In some applications the \( m \)th moment of the distance, rather
than the distance itself, is of interest. As an example, for a collection
of nucleons interacting via simple harmonic oscillator potentials,
\( \left\langle s^{2}\right\rangle  \) may be of interest rather
than \( \left\langle s\right\rangle  \) itself. We then calculate
the \( m \)th moment \( \left\langle s^{m}\right\rangle  \) for
the case of a spherical \( n \)-ball of uniform density, where \begin{equation}
\left\langle s^{m}\right\rangle =\int _{0}^{2R}s^{m}P_{n}(s)\, ds.
\end{equation}
 Using Eq.~(\ref{eq_new_integral}), the \( m \)th moment \( \left\langle s^{m}\right\rangle  \)
has the general form \begin{equation}
\label{eq_moment_uniform_tu}
\left\langle s^{m}\right\rangle =2^{n+m}\left( \frac{n}{n+m}\right) \frac{B\left( \frac{n}{2}+\frac{1}{2},\frac{n}{2}+\frac{1}{2}+\frac{m}{2}\right) }{B\left( \frac{n}{2}+\frac{1}{2},\frac{1}{2}\right) }R^{m},
\end{equation}
 where \( B\left( p,q\right)  \) is the beta function, and \( m\geq -\left( n-1\right)  \). 

Additionally, \( \left\langle s^{m}\right\rangle  \) can be evaluated
for a spherical space having a Gaussian density distribution where
\( \rho \propto e^{-r^{2}/2\sigma ^{2}} \). From Eq.~(\ref{eq_n_pdf_gaussian})
we find,\begin{equation}
\label{eq_moment_gaussian}
\left\langle s^{m}\right\rangle =\lim _{R\rightarrow \infty }\int _{0}^{2R}s^{m}P_{n}(s)ds=(2\sigma )^{m}\frac{\Gamma \left( \frac{n+m}{2}\right) }{\Gamma \left( \frac{n}{2}\right) }.
\end{equation}

In some applications (such as in nuclear physics) involving low-energy
interactions among nucleons the lower limit (zero) should be replaced
by the hard-core radius \( r_{c}\cong 0.5\times 10^{-13} \) cm~\cite{Bohr}.
In such cases the expressions for \( P_{n}(s) \) and \( \left\langle s^{m}\right\rangle  \)
can be expressed as follows:
\begin{equation}
\label{eq_new_200}
P_{n}(s)=\frac{s^{n-1}\int _{\frac{s}{2}}^{R}\left( R^{2}-x^{2}\right) ^{\frac{n-1}{2}}\, dx}{C(2R,0,n)-C(r_{c},0,n)}
\end{equation}
 and\begin{equation}
\label{eq_new_201}
\left\langle s^{m}\right\rangle =\int _{r_{c}}^{2R}s^{m}P_{n}(s)\, ds=\frac{H(R,r_{c},m,n)}{H(R,r_{c},0,n)},
\end{equation}
where\[
C\left( a,m,n\right) =\int _{0}^{a}s^{m+n-1}\, ds\int _{s/2}^{R}\left( R^{2}-x^{2}\right) ^{\left( n-1\right) /2}\, dx,\]
 and \begin{eqnarray}
H\left( R,r_{c},m,n\right)  & = & \frac{(2R)^{n+m}}{n+m}\left[ B\left( \frac{n}{2}+\frac{1}{2},\frac{n}{2}+\frac{1}{2}+\frac{m}{2}\right) -B_{x}\left( \frac{n}{2}+\frac{1}{2},\frac{n}{2}+\frac{1}{2}+\frac{m}{2}\right) \right] \nonumber \\
 &  & -\frac{r_{c}^{n+m}}{n+m}\left[ B\left( \frac{1}{2},\frac{n}{2}+\frac{1}{2}\right) -B_{x}\left( \frac{1}{2},\frac{n}{2}+\frac{1}{2}\right) \right] ,\label{eq_moment_bohr} 
\end{eqnarray}
with \( x=r_{c}^{2}/4R^{2} \), and \( m \) an integer. We note that
Eqs. (\ref{eq_new_200}) and (\ref{eq_new_201}) are the first known
analytical results for \(P_{n}(s)\) and \(\left\langle s^{m}\right\rangle\) 
to incorporate the hard-core radius \( r_{c} \).

\subsection{Neutrino-Pair Exchange Interactions}

A second example of interest is the \( \nu \bar{\nu } \)-exchange
(neutrino-pair exchange) contribution 
to the self energy of a nucleus or a neutron star. For two point masses
the \( 2 \)-body potential energy 
is given by~\cite{Fischbach,Feinberg_1,Feinberg_2,Hsu,wood_001,wood_002,wood_003}
\begin{equation}
V_{\nu \bar{\nu }}\left( \left| \vec{r}_{i}-\vec{r}_{j}\right| \right) =\frac{G_{F}^{2}\, a_{i}a_{j}}{4\pi ^{3}\left| \vec{r}_{i}-\vec{r}_{j}\right| ^{5}},
\end{equation}
 where \( a_{i} \) and \( a_{j} \) are coupling constants which
characterize the strength of the neutrino coupling to fermions \( i \)
and \( j \) (\( i \), \( j \) = electron, proton, or neutron).
In the standard model~\cite{data_group},\begin{eqnarray*}
a_{e} & = & \frac{1}{2}+2\sin ^{2}\theta _{W}=0.964\\
a_{p} & = & \frac{1}{2}-2\sin ^{2}\theta _{W}=0.036\\
a_{n} & = & -\frac{1}{2}.
\end{eqnarray*}
 As an example, we find for the case of a uniform density distribution of radius \(R\) containing 
\( N\) neutrons,
\begin{equation}
\label{eq_moment_hard_core}
W_{3}=\frac{N(N-1)}{2}\left( \frac{3}{2r_{c}^{2}R^{3}}-\frac{9}{4r_{c}R^{4}}+\frac{9}{8R^{5}}-\frac{3r_{c}}{16R^{6}}\right) \frac{G_{F}^{2}}{4\pi ^{3}},
\end{equation}
 where \( r_{c} \) is the hard-core radius. The analogous result
for a Gaussian density distribution is \begin{equation}
W_{3}=\left[ \frac{1}{r_{c}^{2}}e^{-r_{c}^{2}/4\sigma ^{2}}-\frac{\Gamma \left( 0,\, r_{c}^{2}/4\sigma ^{2}\right) }{4\sigma ^{2}}\right] \frac{N(N-1)G_{F}^{2}}{32\sigma ^{3}\pi ^{7/2}},
\end{equation}
where \( \Gamma (a,b) \) is the incomplete gamma function.

\subsection{Neutron Star Models}

Another application of current interest is the self-energy of neutron
star arising from the exchange of \( \nu \bar{\nu } \) pairs~\cite{wood_001,wood_002,wood_003}.
Here we evaluate the probability density functions in \( 3 \) dimensions
for neutron stars with a multiple-shell density distribution,
which is what is typically assumed in neutron star models~\cite{Pines,Shapiro}.
For illustrative purposes, we discuss a spherically symmetric model
with \( 2 \) spherical shells, 
each of uniform density,
where for simplicity we assume shells
of equal thickness. Some other multiple-shell models and their \( n \)-dimensional
probability density functions can be found in~\cite{sjtu}.

For a \( 2 \)-shell model with a uniform density in each shell define
\( \rho =\rho _{1} \) for \( 0\leq r\leq R/2 \) and \( \rho =\rho _{2} \)
for \( R/2\leq r\leq R \), where \( \rho _{1} \) and \( \rho _{2} \)
are arbitrary constants and \( r \) is measured from the center of
the neutron star. Using the preceding formalism we can show that \( P_{3}\left( s\right)  \)
has \( 4 \) different functional forms specified by \( 4 \) regions:

\begin{enumerate}
\item \( 0\leq s\leq \frac{1}{2}R: \)\begin{equation}
P_{3}(s)=\frac{24(\rho _{1}^{2}+7\rho _{2}^{2})s^{2}}{\left( \rho _{1}+7\rho _{2}\right) ^{2}R^{3}}-\frac{36(\rho _{1}^{2}-2\rho _{1}\rho _{2}+5\rho _{2}^{2})s^{3}}{\left( \rho _{1}+7\rho _{2}\right) ^{2}R^{4}}+\frac{12(\rho _{1}^{2}-2\rho _{1}\rho _{2}+2\rho _{2}^{2})s^{5}}{\left( \rho _{1}+7\rho _{2}\right) ^{2}R^{6}},
\end{equation}

\item \( \frac{1}{2}R\leq s\leq R: \)\begin{eqnarray}
P_{3}(s) & = & -\frac{81(\rho _{1}-\rho _{2})\rho _{2}s}{2\left( \rho _{1}+7\rho _{2}\right) ^{2}R^{2}}+\frac{24\rho _{1}s^{2}}{\left( \rho _{1}+7\rho _{2}\right) R^{3}}-\frac{36\rho _{1}(\rho _{1}+3\rho _{2})s^{3}}{\left( \rho _{1}+7\rho _{2}\right) ^{2}R^{4}}\nonumber \\
 &  & +\frac{12\rho _{1}^{2}s^{5}}{\left( \rho _{1}+7\rho _{2}\right) ^{2}R^{6}},
\end{eqnarray}

\item \( R\leq s\leq \frac{3}{2}R: \)\begin{eqnarray}
P_{3}(s) & = & -\frac{81(\rho _{1}-\rho _{2})\rho _{2}s}{2\left( \rho _{1}+7\rho _{2}\right) ^{2}R^{2}}+\frac{24(9\rho _{1}-\rho _{2})\rho _{2}s^{2}}{\left( \rho _{1}+7\rho _{2}\right) ^{2}R^{3}}-\frac{36(5\rho _{1}-\rho _{2})\rho _{2}s^{3}}{\left( \rho _{1}+7\rho _{2}\right) ^{2}R^{4}}\nonumber \\
 &  & +\frac{12(2\rho _{1}-\rho _{2})\rho _{2}s^{5}}{\left( \rho _{1}+7\rho _{2}\right) ^{2}R^{6}},
\end{eqnarray}

\item \( \frac{3}{2}R\leq s\leq 2R: \)\begin{equation}
P_{3}(s)=\frac{192\rho _{2}^{2}s^{2}}{\left( \rho _{1}+7\rho _{2}\right) ^{2}R^{3}}-\frac{144\rho _{2}^{2}s^{3}}{\left( \rho _{1}+7\rho _{2}\right) ^{2}R^{4}}+\frac{12\rho _{2}^{2}s^{5}}{\left( \rho _{1}+7\rho _{2}\right) ^{2}R^{6}}.
\end{equation}
 
\end{enumerate}
We observe that the probability density functions defined in adjacent
regions are continuous across the boundaries separating the regions.

\subsection{RIPS: A New Test for Random Number Generators in \protect\( n\protect \)-Dimensions}

Another interesting application of the present work is as a new test
of random number generators in \( n \)-dimensions. 
Our statistical method follows 
by applying the probability density functions for the random distance
distribution to evaluate the expectation values of \( \vec{r}_{12}\cdot \vec{r}_{23} \)
in \( n \)-dimensions, where \( \vec{r}_{12}=\vec{r}_{2}-\vec{r}_{1} \),
\( \vec{r}_{23}=\vec{r}_{3}-\vec{r}_{2} \), and \( \vec{r}_{1} \),
\( \vec{r}_{2} \), and \( \vec{r}_{3} \) are three random points
independently sampled from a spherical \( n \)-ball. The quantity
\( \left\langle \vec{r}_{12}\cdot \vec{r}_{23}\right\rangle _{n} \)
is one of the geometric probability constants~\cite{sjtu} and has
applications in many areas of physics~\cite{Fischbach,sjtu}. 
It follows from the preceding formalism that 
for a spherical \( n \)-ball of radius \( R \) with a uniform
density distribution, 
\begin{equation}
\left\langle \vec{r}_{12}\cdot \vec{r}_{23}\right\rangle _{n}=-\frac{n}{n+2}R^{2}.
\end{equation}
 Note that \begin{equation}
\lim _{n\rightarrow \infty }\left\langle \vec{r}_{12}\cdot \vec{r}_{23}\right\rangle _{n}=-R^{2}.
\end{equation}
 In a spherical Gaussian space where \( \rho \propto e^{-r^{2}/2\sigma ^{2}} \)
and \( R\rightarrow \infty  \), \begin{equation}
\left\langle \vec{r}_{12}\cdot \vec{r}_{23}\right\rangle _{n}=-n\sigma ^{2}.
\end{equation}
We refer to this test as RIPS (Random Inner Product in a Sphere).

We now apply the RIPS test to check three popular
random number generators frequently used in Monte Carlo simulations~\cite{Knuth,Numerical,vatt,holian,weyl_002}.
The random number generators tested are:
\begin{enumerate} 
\item RAN\( 0 \)~\cite{Numerical}
which is a linear congruential generator and uses the following algorithm
\begin{equation}
I_{n}=16807\times I_{n-1}\quad \mathrm{mod}\quad m=2^{31}-1.
\end{equation}
\item R\( 31 \)~\cite{vatt} which uses the generalized feedback
shift register (GFSR) method
\begin{equation}
x_{n}=x_{n-p}\oplus x_{n-q,}
\end{equation}
 where \( p=31 \), \( q=3 \), and \( \oplus  \) is the bitwise
exclusive OR operator.
\item NWS~\cite{holian,weyl_002} which uses
the nested Weyl sequence
\begin{equation}
Y_{n}=\left\{ n\left\{ n\alpha \right\} \right\} ,
\end{equation}
 where \( \left\{ y\right\}  \) is the fractional part of \( y \)
and \( \alpha  \) is an irrational number. 
\end{enumerate}
The results are shown
in Table~\ref{table_test}. We have checked several initialization methods and the results
are not affected. 
We note that the NWS generator using the nested Weyl sequence method fails to pass our
randomness tests in all dimensions selected.
Hence caution should be exercised in using this particular random
number generator for Monte Carlo simulations,
especially in molecular dynamics simulations.
More statistical tests for a variety of random number generators
including the PSLQ algorithm using the binary digits of \( \pi  \)~\cite{pi_bailey},
and other newly proposed computational schemes for \( n \)-dimensions,
will be discussed in greater detail in Ref.~\cite{sjtu2}.

\setlongtables
\begin{table}[t]
\caption{Comparison of random number generators (RNG) with the exact results derived in this paper for 
\protect\( \left\langle \vec{r}_{12}\cdot \vec{r}_{23}\right\rangle _{n}\protect \).
The number of simulation samples in each case is \protect\( N=10^{6}\protect \).
We note that the NWS generator fails to pass the new randomness test
in all dimensions that we selected. See text for further discussion.\label{table_test}}
\vspace{0.1 true in}
\begin{tabular}{|c|c|c|c|c|c|c|}
\hline 
 RNG      &\( n=3 \)              &Result&\( n=5 \)              &Result&\( n=10 \)             &Result\\
\hline
RAN\( 0 \)&\( -.60037\pm .00065 \)&Pass     &\( -.71483\pm .00059 \)&Pass  &\( -.83310\pm .00047 \)&Pass  \\
R\( 31 \) &\( -.60028\pm .00066 \)&Pass     &\( -.71468\pm .00058 \)&Pass  &\( -.83229\pm .00048 \)&Pass  \\
NWS       &\( -.64119\pm .00071 \)&\bf{Fail}&\( -.75427\pm .00057 \)&\bf{Fail}&\( -.85016\pm .00049 \)&
\bf{Fail}  \\
\hline
Exact     &\( -.60000 \)          &      &\( -.71429 \)          &      &\( -.83333 \)          &      \\
\hline
\end{tabular}
\end{table}

The geometric probability constants,
\( \left\langle \vec{r}_{12}\cdot \vec{r}_{23}\right\rangle _{n}\),
can be added to
the family of various computational tests for random number generators.
They can then serve to investigate the quality of random number generators
for questions of randomness,
 especially in higher dimensions (\( n>3 \)) 
where few results are currently available. 
The possibility that some random number generators such as NWS pass other tests
but not ours, may indicate that the properties of random points in a sphere
provide a more sensitive test of randomness than is otherwise available.
These and other issues will be discussed in more detail in Ref.~\cite{sjtu2}.

\section{Conclusions}

A formalism has been presented in this paper for evaluating the analytical
probability density function of the random distance distribution for
a spherical \( n \)-ball with an arbitrary density distribution.
We show that the random distance distribution technique can reduce
otherwise difficult calculations from the complexity of \( 2n \)-dimensional integrals 
to just a \( 1 \)-dimensional integral, even when the nucleon-nucleon
hard-core radius \( r_{c} \) is included. 
Our formalism has applications to the currently active area of 
research surrounding string-inspired theories of higher dimensional physics, 
and has numerous potential applications to other fields as well. 
Specifically the results presented here are
of interest in the context of recent work on the modifications to
the Newtonian inverse-square law arising from the existence of extra
spatial dimensions~\cite{ArKani}. We have also presented a new computational
method to test random number generators in \( n \)-dimensions used
in Monte Carlo simulations. 

\section{Acknowledgments}

The authors wish to thank Chinh Le, Michelle Parry, David Schleef,
Dave Seaman, Christopher Tong, and A. W. Overhauser for helpful discussions
and the Purdue University Computing Center for computing support.
This work was supported in part by the U.S. Department of Energy under
Contract No. DE-AC02-76ER01428.

\bibliographystyle{apsrev}

\end{document}